# Electrochemical Sensing of Lead in Drinking Water Using MWCNTs and β-Cyclodextrin

*Arif Ul Alam,[1,2] Matiar M. R. Howlader,[1*] Nan-Xing Hu,[2] and M. Jamal Deen[1*]*

[1]Department of Electrical and Computer Engineering, McMaster University, 1280 Main Street West, Hamilton, ON, L8S 4K1, Canada

[2]Advanced Materials Laboratory, Xerox Research Centre of Canada, 2660 Speakman Drive, Mississauga, ON, L5K 2L1, Canada

**ABSTRACT**:

Heavy metal pollution is a severe environmental problem affecting many water resources. The non-biodegradable nature of the heavy metals such as lead (Pb) causes severe human health issues, so their cost-effective, sensitive and rapid detection is needed. In this work, we describe a simple, facile and low cost modifications of multiwalled carbon nanotubes (MWCNT) and β-cyclodextrin (βCD) through non-covalent/physical (Phys) and a covalent Steglich esterification (SE) approaches. The Phys modification approach resulted Pb detection with a limit-of-detection (LoD) of 0.9 ppb, while the SE approach showed an LoD of 2.3 ppb, both of which are well below the WHO Pb concentration guideline of 10 ppb. The MWCNT-βCD (Phys) based electrodes show negligible interference with other common heavy metal ions such as $Cd^{2+}$ and $Zn^{2+}$. The MWCNT-βCD based electrodes were of low-cost owing to their simple synthesis approaches, exhibited good selectivity and reusability. The proposed MWCNT-βCD based electrodes is a promising technology in developing a highly affordable and sensitive electrochemical sensing system of Pb in drinking water.

Accessibility to potable water is increasingly challenging in developing and in some developed countries due to contamination of their source waters with heavy metal ion and other pollutants.[1] Heavy metals such as lead (Pb) are non-biodegradable and widely distributed and its presence in drinking water causes greater risks to human health.[2] The effects of Pb include behavioral disorder and neurodevelopmental problems in children; increased blood pressure and renal dysfunction in adults; and even cancer in kidneys, lung, or brain due to its long-term presence in source and drinking water.[3–7] According to World Health Organization (WHO) guidelines, the maximum acceptable concentration of Pb in drinking water should be 10 µg/L or 10 ppb.[6] However, drinking water authorities such as in Canada are proposing an even stricter limit (e.g., 5 ppb) for Pb in drinking water.[6] Major challenges in implementing the stricter limit include the lack of on-site monitoring techniques, and detecting contaminant levels across distributed water sources. Therefore, a simple, low-cost and easy to use sensor for the detection of a heavy metal such as Pb is necessary to maintain water safety in resource-limited areas.

Conventional analytical techniques such as inductively coupled plasma mass spectrometry and atomic absorption spectroscopy require qualified testing laboratories and trained personnel.[8,9] Recently, electrochemical methods have made considerable progress towards simple, on-site and low-cost detection capabilities to allow adequate time for taking safety measures in case of a contamination.[10] The electrochemical sensors, commonly referred to as "electrodes", are ideal candidates as they can be fabricated with low-cost to detect Pb with higher precision and accuracy. The material system of a sensing electrode is the key ingredient in maximizing overall performance of an electrochemical analysis instrument. Specifically, electrode materials based on carbon nanomaterials,[11] metal nanoparticles,[12] and a number of selectivity-enhancing polymeric or organic materials[13–15] have attracted tremendous research interest for their collective attributes such as high effective surface area, enhanced electron transfer, ability to be miniaturized, and improved selectivity. More notably, nanomaterials based electrodes made up of easily available electrode materials by a facile and scalable synthesis method can be exploited to fabricate low cost electrodes for the analysis of multiple environmental targets such as heavy metals such as lead and pharmaceutical contaminants such as acetaminophen and estrogen in drinking water.[16,17] The ability to detect multiple environmental targets at low cost and in an easy to use manner requires careful selection of sensing materials to provide high sensitivity and selectivity, development of simple modification approaches and user-friendly analysis steps.

Multiwalled Carbon nanotubes (MWCNTs) are excellent electrode material to realize simple, low cost and easy-to-use electrochemical sensors because of their wide electrical potential window, fast electron transfer rate and large surface area.[18] Unmodified MWCNTs in the form of tower, array, and thread structures were used as an electrode material. These unmodified MWCNTs provided considerable simplicity in electrode fabrication compared to that of modified MWCNTs.[19] However, MWCNTs do not disperse well in most common organic and inorganic solvents due to the high Van der Waals force between them that lead to aggregation and bundling.[10] Therefore, non-covalent modification of MWCNTs through $\pi$-$\pi$ conjugation between MWCNTs surface and the organic modifiers such dodecylbenzene sulfonate and Nafion showed better sensing performances.[11] However, non-covalent modifications cannot significantly improve the sensitivity and selectivity of the electrodes since individual nanotubes becomes less interconnected, thus facilitating fewer electron transfer.[11] MWCNTs have also been covalently modified by functionalization with amino groups ($-NH_2$), thiol groups ($-SH$), amino acids ($-NH^{3+}$ and $-COOH$), and Bismuth (Bi) which improved the selectivity and limit of detection.[11] But then the covalent modification approaches require tedious chemical reaction procedures and thus increase fabrication complexity.

Recently, non-covalent modifications of MWCNTs with $\beta$-cyclodextrin ($\beta$CD) molecule have shown enhanced sensing performance due to improved dispersion of MWCNTs in the presence of $\beta$CD.[20,21] The improved sensing performance was due to the inclusion complex formation property of $\beta$CD towards the detection of emerging pharmaceutical contaminants such as acetaminophen and estrogen in drinking water.[20,21] Remarkably, βCD have also shown selective formation of inclusion complexes with Pb ions (Pb$^{2+}$) which can be exploited for the detection Pb$^{2+}$.[15] βCD functionalized gold nanoparticles have also shown selective detection of Pb$^{2+}$.[14] Therefore, in this work, we show that the enhanced electrochemical properties of MWCNTs can be combined with the inclusion capability of βCD through their non-covalent and covalent conjugation towards the sensitive detection of Pb$^{2+}$. We demonstrate that the non-covalently modified MWCNT-βCD shows the best Pb$^{2+}$ sensing performance with at least four times reusability. On the other hand, the covalently modified MWCNT-βCD shows moderate Pb$^{2+}$ sensing performance with at least ten times reusability. Thus, the type of conjugation between MWCNTs and βCD, that is, non-covalently and covalently modified MWCNT-βCD, significantly controls the sensing performance of the electrodes.

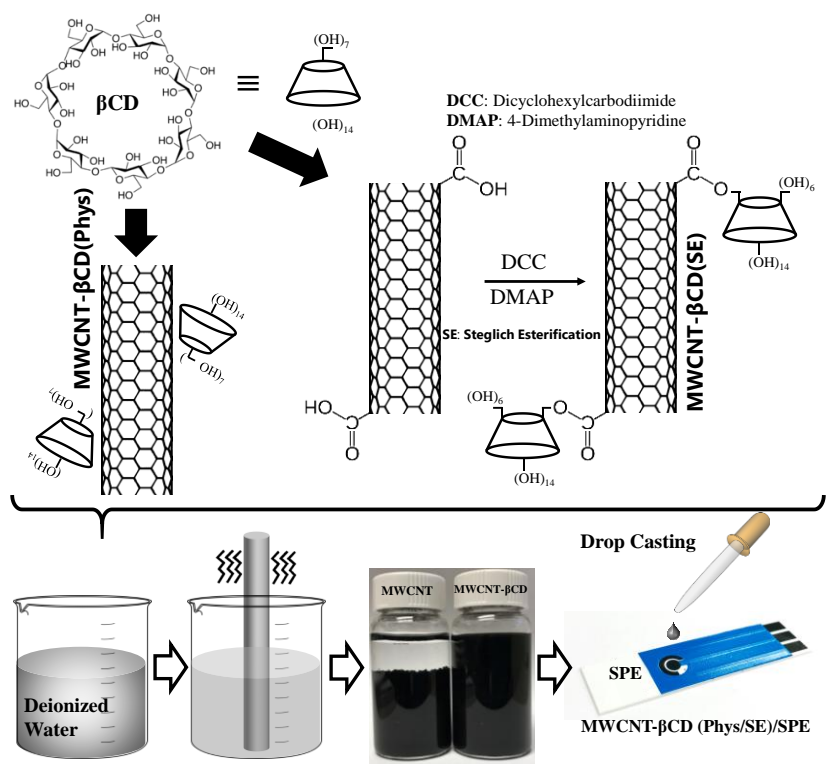

**Scheme 1:** Schematic diagram of the MWCNT-βCD (Phys/SE)/SPE electrode fabrication processes.

The non-covalent modification of MWCNTs and βCD was done by physically mixing them in **D**e**i**onized **W**ater (DIW). Twenty milligram of MWCNTs (purchased from U.S. Research Nanomaterials

Inc., outer diameter (OD): 5−15 nm, length: ~50 µm, and purity: ≥ 95 wt%) was dispersed by using an ultrasonic preprocessor in 10 ml of βCD solution (2 wt%) in DIW to give a 2 mg mL$^{-1}$ stable black suspension (Scheme 1). The resulting suspension will be termed as MWCNT-βCD (Phys) since the modification process is based on physical mixing of MWCNTs and βCD. An aliquot of 10 µL of 2 mg mL$^{-1}$ MWCNT-βCD (Phys) solution was drop-casted on a commercial **S**creen **P**rinted **E**lectrode (SPE) (Zensor, purchased from CH Instruments, with Carbon working and counter electrode, and Ag/AgCl reference electrode). The drop-casted MWCNT-βCD (Phys) solution was dried in air for 20 min at 80 °C in an oven resulting in MWCNT-βCD (Phys)/SPE electrode (Scheme 1).

The covalent modification of MWCNTs and βCD was accomplished by a simple, one-step chemical reaction based on **S**teglich **e**sterification (SE) principle.[17] Briefly, MWCNT-COOH (purchased from U.S. Research Nanomaterials Inc., OD: 5−15 nm, length: ~50 µm, and purity: ≥ 95 wt%) was first dispersed in anhydrous *N,N*-dimethylformamide (DMF) into a reactor with Ar environment. The mixture in the reactor was then sonicated to get a uniform suspension. After that 4-(dimethylamino)pyridine (DMAP) and βCD and dicyclohexylcarbodiimide (DCC) were added into the reaction mixture at 0 °C. The resulting suspension was stirred for several hours at 20 °C to precipitate MWCNT-βCD (SE)), also termed as SE modified MWCNTs and βCD. The MWCNT-βCD (SE) was then filtered and washed repeatedly to remove the DCC, DMAP, and unreacted βCD. The SE process couples MWCNT-COOH and βCD through ester bonds resulting in the chemically or covalently modified MWCNTs with βCD. A uniform, black suspension of the covalently modified MWCNT-βCD (SE) was prepared by dispersing 40 mg of the black MWCNT-βCD (SE) powder into 20 mL of DIW using an ultrasonic preprocessor. Finally, an aliquot of 10µL 2 mg ml$^{-1}$ MWCNT-βCD (SE) solution was drop-casted on SPE and then dried in air for 20 min at 80 °C in an oven resulting in MWCNT-βCD (SE)/SPE electrode (Scheme 1). For comparison, an SPE electrode was also modified by drop casting only MWCNT (i.e., 2mg/mL MWCNT dispersed in DIW using ultrasonic preprocessor) and dried to get MWCNT/SPE.

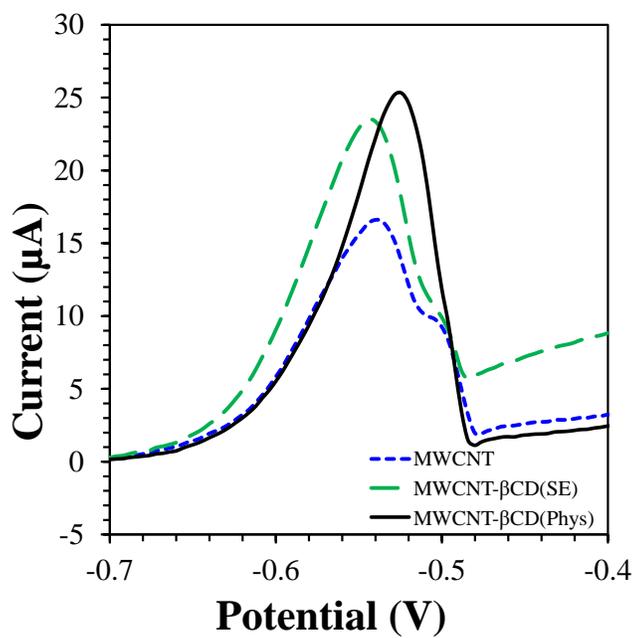

**Figure 1**: DPASV of MWCNT/SPE (blue smaller dashed curve), MWCNT-βCD (SE)/SPE (green larger dashed curve), and MWCNT-βCD (Phys)/SPE (black solid curve) recorded in 0.1 M acetate buffer (pH 5) in the presence of 1 μM (207 ppb) Pb$^{2+}$ ion.

The Pb$^{2+}$ ion sensing was done by **D**ifferential **P**ulse **A**nodic **S**tripping **V**oltammetry (DPASV) method. The test solutions with different Pb$^{2+}$ ion concentrations were made by dissolving appropriate amount of Pb(NO$_3$)$_2$ salt in 0.1 M acetate buffer (pH 5). A miniature USB powered potentiostat (EmStat3 from PalmSens) with PSTrace 5.5 data analyzer software was used in the electrochemical measurements. At the beginning of the electrochemical measurement, a deposition potential of −0.8 V was applied at the working electrode for 600 s under magnetic stirring (250 rpm) in 50 mL solution to reduce Pb$^{2+}$ ions to Pb$^0$. The electrodeposited Pb atoms were then stripped from the working electrodes by anodically sweeping the electrode potential from −0.8 to −0.2 V. The optimized DPASV parameters were 4 mV, 50 mV, 50 ms and 40 mV/s for step potential, pulse amplitude, pulse time and scan rate, respectively. The peak currents were measured with respect to linear base line of the current vs. potential curve.

There are two major approaches to improve the performance of an electrochemical sensor. First, the effective surface area of the working electrode plays a key role in achieving high sensitivity. The effective surface area of the MWCNT-βCD (Phys)/SPE and MWCNT-βCD (SE)/SPE is higher compared to that of MWCNT/SPE.[17] Second, the target-capturing ability of the sensing material system can improve its selectivity and dynamic range. A comparison of the sensing performance of the prepared MWCNT-βCD (Phys)/SPE and MWCNT-βCD (SE)/SPE with that of MWCNT/SPE in the presence of 1 μM (207 ppb) $Pb^{2+}$ is shown in Figure 1. The peak current for MWCNT/SPE was 15.1 μA. The MWCNT-βCD (SE)/SPE showed increased current peak of 19.3 μA, whereas the MWCNT-βCD (Phys)/SPE showed the highest peak current of 24.4 μA. The peak potential of MWCNT-βCD (Phys)/SPE was slightly shifted towards more positive direction and this could be due to modification in the oxidation potential in the presence of βCD.

The enhanced electrochemical sensing of $Pb^{2+}$ ion with MWCNT-βCD (Phys)/SPE can be attributed to higher amount of $Pb^0$ deposition during application of the deposition potential or −0.8 V. To confirm this phenomenon, we performed **S**canning **E**lectron **M**icroscopy (SEM) observations of the three different electrodes (MWCNT/SPE, MWCNT-βCD (SE)/SPE and MWCNT-βCD (Phys)/SPE) taken out of the electrochemical cell just after the deposition step. These electrodes were used since performing the stripping step will cause anodization induced-stripping of the $Pb^0$. Figure 2 shows the corresponding SEM images of the modified electrodes with low (on top, x200) and high (on bottom, x1000) magnifications. The SEM images were taken using a JEOL 7100F SEM microscope with an acceleration voltage of 15 kV and working distance of 6 mm. Figure 2(a) shows that the MWCNT/SPE has dendritic structures of the electrodeposited $Pb^0$. On the other hand, as shown in Figure 2(b), the MWCNT-βCD (SE)/SPE surface has island structures of the electrodeposited $Pb^0$, which covers more surface area to that of MWCNT/SPE. However, as shown in Figure 2(c), the MWCNT-βCD (Phys)/SPE surface is covered with leaf-life structure of the accumulated $Pb^0$. All the low and high magnification SEM images are scaled equally. Therefore, Figure 2 shows that the MWCNT-βCD (Phys)/SPE has the highest amount of surface coverage by the electrodeposited $Pb^0$. The highest surface coverage of $Pb^0$ resulted in the largest peak current due to anodic oxidation stripping from the MWCNT-βCD (Phys)/SPE electrode (Figure 1).

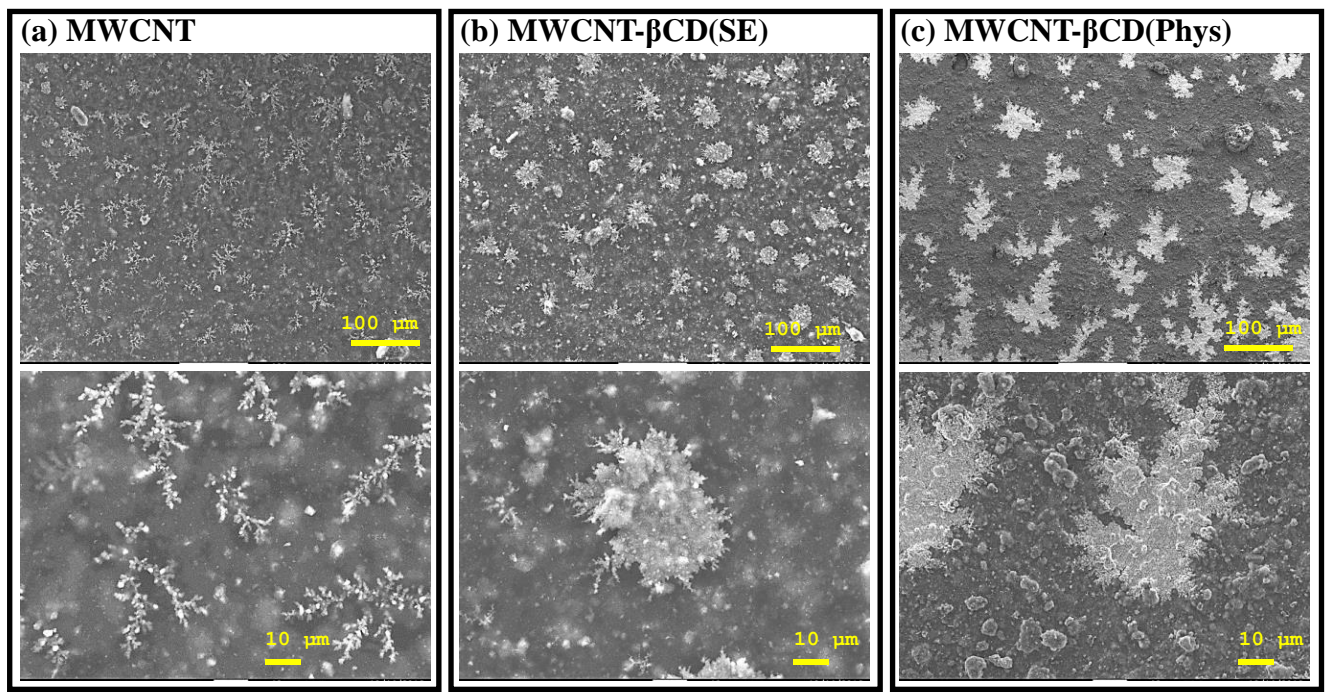

**Figure 2**: SEM images of the (a) MWCNT/SPE, (b) MWCNT-βCD (SE)/SPE, and (c) MWCNT-βCD (Phys)/SPE with low (upper images) and high (lower images) magnifications after electrodeposition steps (at −0.8 V) of the corresponding electrodes in the presence of 1 μM (207 ppb) $Pb^{2+}$ ion.

The difference in the surface coverage by Pb in the three types of electrodes could be related to the amount of βCD present with the MWCNTs. We demonstrated before that the amount of βCD is highest in MWCNT-βC (Phys) compared to that of MWCNT-βCD (SE).[17] In fact, there are 1:2 weight ratio of βCD:MWCNTs present in the MWCNT-βCD (Phys) suspension through physical/non-covalent mixing. On the contrary, MWCNT-βC (SE) has much smaller amount of βCD since the chemical attachment of βCD with MWCNT (i.e., the Steglich esterification) depends on the amount of –COOH groups present in the precursor MWCNT-COOH material. It was found that MWCNT-βC (SE) has only ~1:10 weight ratio of βCD:MWCNTs.[17] Thus, it is confirmed that MWCNT-βC (Phys)/SPE electrode has the highest amount of βCD which are physically/non-covalently attached with MWCNTs. Hence, the increased amount of βCD in MWCNT-βCD (Phys)/SPE facilitated improved electrochemical sensing of $Pb^{2+}$.

The improved electrochemical sensing performance of MWCNT-βCD (Phys)/SPE compared to that of MWCNT/SPE and MWCNT-βCD (SE)/SPE could also be attributed to the formation of the inclusion complexes between βCD and $Pb^{2+}$. The βCD molecule has a porous structure with an average diameter of 18–20 nm which is capable to form host-guest interaction based inclusion complexes with other molecules. A number of researches had investigated its host-guest interaction properties with phenolic organic molecules and steroid hormones.[16,17] These organic molecules and hormones are attracted to the hydrophobic inner core of the βCD molecule and facilitates redox-based electrochemical sensing of the guest molecules. The βCD also showed affinity towards metal ions. For example, βCD functionalized gold-iron nanoparticles was used for selective and sensitive colorimetric sensing of $Cr^{6+}$ ion. In another study,[22] carboxymethyl-β-cyclodextrin polymer modified $Fe_3O_4$ nanoparticles (CDpoly-MNPs) were utilized for selective removal of $Pb^{2+}$, $Cd^{2+}$, and $Ni^{2+}$ ions from water. The CDpoly-MNPs showed preferential adsorption towards $Pb^{2+}$ ions with an affinity order of $Pb^{2+} \gg Cd^{2+} \gg Ni^{2+}$. Moreover, βCD functionalized gold nanoparticles were used for the selective detection of $Pb^{2+}$ ions from aqueous solution. The electrochemical evidence of the interaction between the $Pb^{2+}$ ions and the βCD was also described where the formation constants of surface inclusion complexes between $Pb^{2+}$ and βCD were electrochemically measured ($727.5\pm20.2$ $M^{-1}$) using a carbon paste electrode.[15,23] Therefore, in this study, the inclusion complex formation of βCD has been combined with the superior electrochemical properties of MWCNTs towards sensing of $Pb^{2+}$ ion.

The linear detection range of MWCNT-βCD (Phys)/SPE electrode for the detection of Pb2+ ion was measured to determine its feasibility for use in drinking water monitoring. Figure 3 shows the DPASV curves and the corresponding linear calibration curve in the presence of 0.015−20 µM (3.1–103.3 ppb) $Pb^{2+}$ ion. The optimized DPASV conditions used in these measurements are similar to that of Figure 1.

Figure 3(a) shows that the current peak for lowest Pb$^{2+}$ ion concentration is located at −0.6 V. The peak current slightly shift towards more anodic value with the increased concentration of Pb$^{2+}$ ion. This peak shifting could be due to diffusional control reaction influenced by concentration difference.[24] Figure 3(b) shows the calibration curves for the concentration ranges of 0.015–20 μM (3.1-103.3 ppb). The error bars for each data point was calculated by taking six measurements using newly prepared sensors each time and by multiplying Student's $t$-distribution $k$ value of 2.57 to get 95% confidence interval. The linear regression equations can be expressed as $I_p = 0.098\ C + 0.0266$, where $I_p$ is the peak current (in μA) and $C$ is the Pb$^{2+}$ concentration (in ppb).

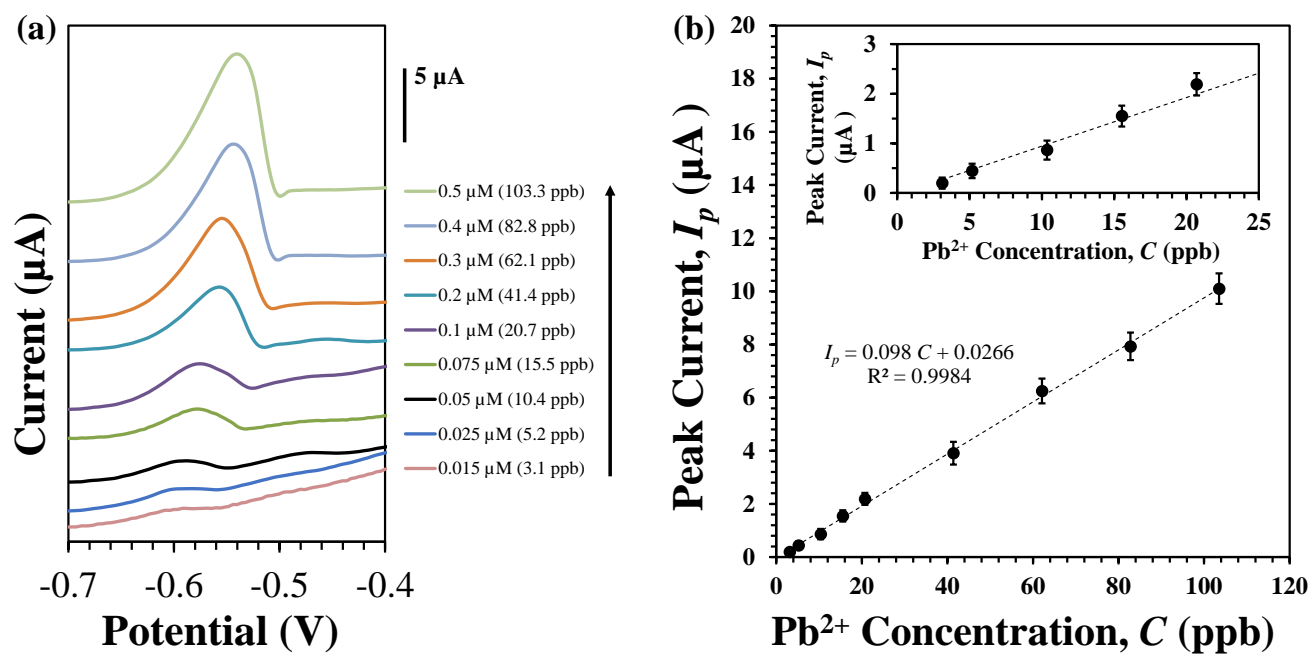

**Figure 3:** (a) DPASV of MWCNT-βCD (Phys)/SPE in 0.1 M acetate buffer (pH 5) in the presence of 0.025−20 μM (3.1−103.3 ppb) Pb$^{2+}$. (b) Calibration curve (i.e., peak current vs Pb$^{2+}$ concentration) for MWCNT-βCD (Phys)/SPE. All experiments were done with a deposition potential of −0.8 V for 600 s under magnetic stirring before measurements.

The sensitivity of the MWCNT-βCD (Phys)/SPE electrode is 98 nA/ppb. The limit of detection (LoD), defined as LoD = 3$s$/m (where, $s$ is the standard deviation of the blank solution (0.03 μA), and $m$ is the slope of the calibration curve), was estimated to be 0.9 ppb. Therefore, the linear range and the LoD value of the MWCNT-βCD (Phys)/SPE sensor to detect lead concentration in drinking water are much better than those specified by the WHO guideline. The LoD value is compared with other recent reports for the detection of $Pb^{2+}$ using electrodes based on MWCNTs, as shown in Table S1.[25–30] It is observed that the MWCNT-βCD (Phys)/SPE sensor has comparable linear range and one of the lowest limit of detection (LoD) for the detection of $Pb^{2+}$. For comparison, the linear range and calibration curves of MWCNT-βCD (SE)/SPE electrode was also measured and are shown in Figure S1. The MWCNT-βCD (SE)/SPE electrode showed a linear detection range of 6.2 – 103.5 ppb with an LoD of 2.3 ppb. Thus, the sensitivity of the MWCNT-βCD (SE)/SPE electrode is 38.6 nA/ppb, which is ~2.5 times lower than that of MWCNT-βCD (Phys)/SPE. However, the MWCNT-βCD (SE)/SPE electrode showed better performance in terms of stability.

The stability of the MWCNT-βCD (Phys)/SPE and MWCNT-βCD (SE)/SPE electrodes were investigated to examine their repeatability and reusability. The two types of electrode were repeatedly reused in similar and different concentrations of $Pb^{2+}$ solutions. Before reusing the electrodes, they were electrochemically cleaned/regenerated by applying 0.31 V under magnetic stirring in a blank solution for 300 s to desorb the Pb molecules from the electrode surface, as shown in Figure S2. The cleaning procedure was found to be more effective and faster when magnetic stirring was used, as shown in Figure S2(a) and S2(b). Figure S2(c) shows the DPASV response of a newly prepared MWCNT-βCD (Phys)/SPE electrode with 207 ppb of $Pb^{2+}$ (curve ①), which was then electrochemically cleaned to give a DPASV curve in blank solution with no peak current (curve ②), signifying successful regeneration of the electrode. After that, the same electrode was used to DPASV measurement of 5.2 ppb of $Pb^{2+}$ with peak current of 0.4 µA (curve ③), which is within the error bar range of the calibration line of MWCNT-βCD (Phys)/SPE. However, the MWCNT-βCD (Phys)/SPE electrode showed stable sensing performance when reused/regenerated for four consecutive measurements of 207 ppb of $Pb^{2+}$ solutions.

On the other hand, the MWCNT-βCD (Phys)/SPE electrode performed at least six consecutive reuse/regeneration cycles with negligible degradation in sensing performance when low concentrations of $Pb^{2+}$ solution (in the range of 5−20 ppb) were used. Therefore, the MWCNT-βCD (Phys)/SPE electrode tends to degrade faster when measuring higher concentration of $Pb^{2+}$. This could be due to dissolution of βCD molecule from the electrode surface when increased amounts of $Pb^0$ molecules adsorb which ultimately leads to delamination and disintegration of the MWCNTs network during the electrode regeneration step. In contrast, the MWCNT-βCD (SE)/SPE electrode showed reusability for at least 10 reuse/regeneration cycles at both low and high concentrations of $Pb^{2+}$. The better reusability of the MWCNT-βCD (SE)/SPE electrode can be attributed to the chemical/covalent attachment of βCD with MWCNTs which is stronger than that of the MWCNT-βCD (Phys) that provides better electrode surface adhesion during cleaning/regeneration steps. Consequently, the MWCNT-βCD (SE)/SPE can provide better reusability when measuring higher levels of $Pb^{2+}$, and the MWCNT-βCD (Phys)/SPE can offer better sensitivity as well as reusability when lower levels of $Pb^{2+}$ is measured.

The MWCNT-βCD (Phys)/SPE and MWCNT-βCD (SE)/SPE electrodes are low cost and can be used utilized in resource limited areas. The low cost of fabrication of these electrodes is due to the simple MWCNTs modification processes (covalent/non-covalent), use of simple equipment and low-cost commercial carbon SPE electrodes. Furthermore, apart from the electrochemical cleaning/regeneration procedure described above, the SPE electrode can also be reused by simply wiping off the MWCNT-βCD from its surface using Kim Wipes and cleaning with solvents like isopropyl alcohol followed by drop-casting another aliquot of 10 µL of MWCNT-βCD solution. The SPE electrode showed negligible degradation in sensing performance over at least 50 cycles of cleaning and remaking of the electrodes.

The reusability of the SPE electrode provides further economic benefits to reduce the overall cost to run a single experiment. Table S2 shows the breakdown of materials cost (in US$) to fabricate a new MWCNT-βCD (Phys)/SPE or MWCNT-βCD (SE)/SPE electrode and to run a single experiment to

highlight the cost-effectiveness of the proposed $Pb^{2+}$ sensor. Table S2 shows that the materials cost a 10 µL drop of MWCNT-βCD (Phys/SE) required to fabricate a single sensor is only $0.0026 which is only 0.21% compared to the price of a single SPE ($1.25). Additionally, the prices included on Table S2 can be reduced more since it lists the retail prices. Also, if a MWCNT-βCD (Phys/SE)/SPE sensor is cleaned/regenerated for 10 times, then the cost of each measurement will be much smaller. Therefore, the MWCNT-βCD (Phys/SE)/SPE electrode can be fabricated with extremely low cost. The fabrication costs can be further reduced by fabricating the SPE electrode also instead of using the commercial one.

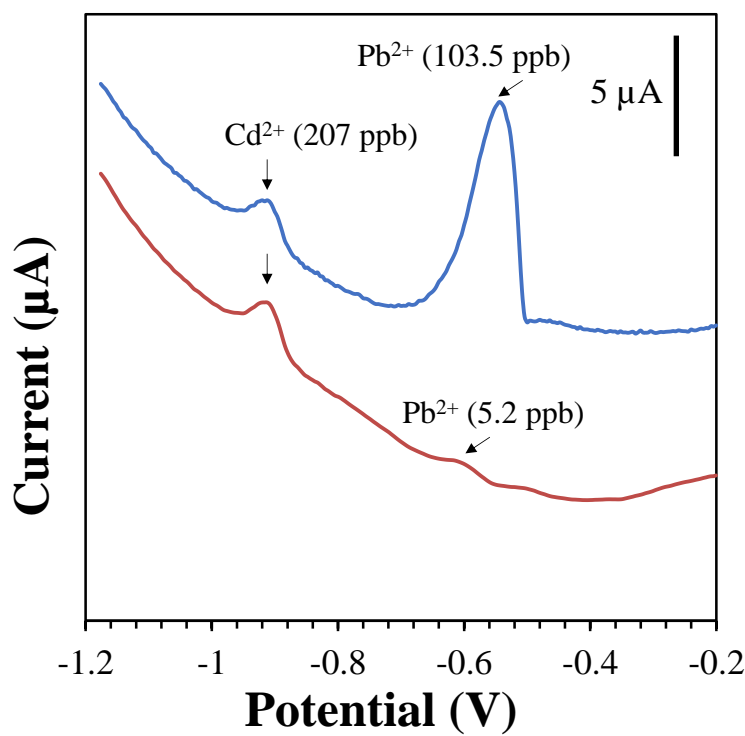

**Figure 4:** DPASV of MWCNT-βCD (Phys)/SPE with 5.2 ppb and 103.5 ppb of $Pb^{2+}$ in the presence of 207 ppb of $Zn^{2+}$ and $Cd^{2+}$ in 0.1 M acetate buffer (pH = 5).

An important performance parameter of an electrochemical heavy metal sensor is its selectivity in the presence of other interfering ions in the solution. The selectivity of the MWCNT-βCD (Phys)/SPE electrode with 103.5 ppb of $Pb^{2+}$ was measured in the presence of 207 ppb of Cadmium ($Cd^{2+}$) and Zinc ($Zn^{2+}$) ions (Figure 4). The $Cd^{2+}$ peak was clearly separated from the $Pb^{2+}$ peak. Also, the $Pb^{2+}$ peak showed negligible change in the peak current. However, the $Zn^{2+}$ peak was not observed since the oxidation potential of $Zn^{2+}$ is lower than $-1.2$ V and the deposition potential ($-0.8$ V) did not cause any deposition of $Zn^0$ molecule on the electrode surface. The peak current for 207 ppb of $Cd^{2+}$ was very small compared to that of $Pb^{2+}$. Therefore, the MWCNT-βCD (Phys)/SPE electrode showed more selectivity towards $Pb^{2+}$ ion.

In this research work, we developed low-cost and easy-to-use electrochemical sensors for the determination of $Pb^{2+}$ using non-covalent/physical and a one-step covalent modifications (Steglich esterification) of MWCNT with βCD. The modified MWCNT-βCD electrodes demonstrated excellent sensing performance for the detection of $Pb^{2+}$ with limits-of-detection of 0.9 and 2.3 ppb for the non-covalent/physical and covalent modifications, respectively. The physically modified MWCNT-βCD based electrode exhibited the highest sensitivity compared to that of MWCNT and covalently modified MWCNT-βCD. The electrodes fabrications and materials were of extremely low cost which was calculated to be only $2.64 per 1000 MWCNT-βCD modifications. The final cost of each measurement could be further reduced because of the reusability of the MWCNT-βCD based electrodes for at least four times. The sensors also showed good selectivity in the presence of other interfering heavy metal ions such as $Zn^{2+}$ and $Cd^{2+}$ with very high concentrations. The developed sensor may be used in low-cost and point-of-care sensing systems for drinking water quality monitoring.

# AUTHOR INFORMATION

## Corresponding Author

*Email: jamal@mcmaster.ca, mrhowlader@ece.mcmaster.ca

## Author Contributions

The manuscript was written through contributions of all authors. All authors have given approval to the final version of the manuscript.

## Notes

The authors declare no competing financial interest.

# ACKNOWLEDGEMENTS

The authors acknowledge the valuable feedback and suggestions given by the research group members of Dr. M Jamal Deen during the research and writing of the manuscript. This research is supported by Discovery Grants from the Natural Science and Engineering Research Council of Canada, an infrastructure grant from the Canada Foundation for Innovation, an Ontario Research Fund for Research Excellence Funding Grant, a FedDev of Southern Ontario grant, the Canada Research Chair program, NSERC RES'EAU strategic network and the NCE IC-IMPACTS.

**TABLE OF CONTENT GRAPHIC (Width and Length Ratio: 3.6 × 8.4)**

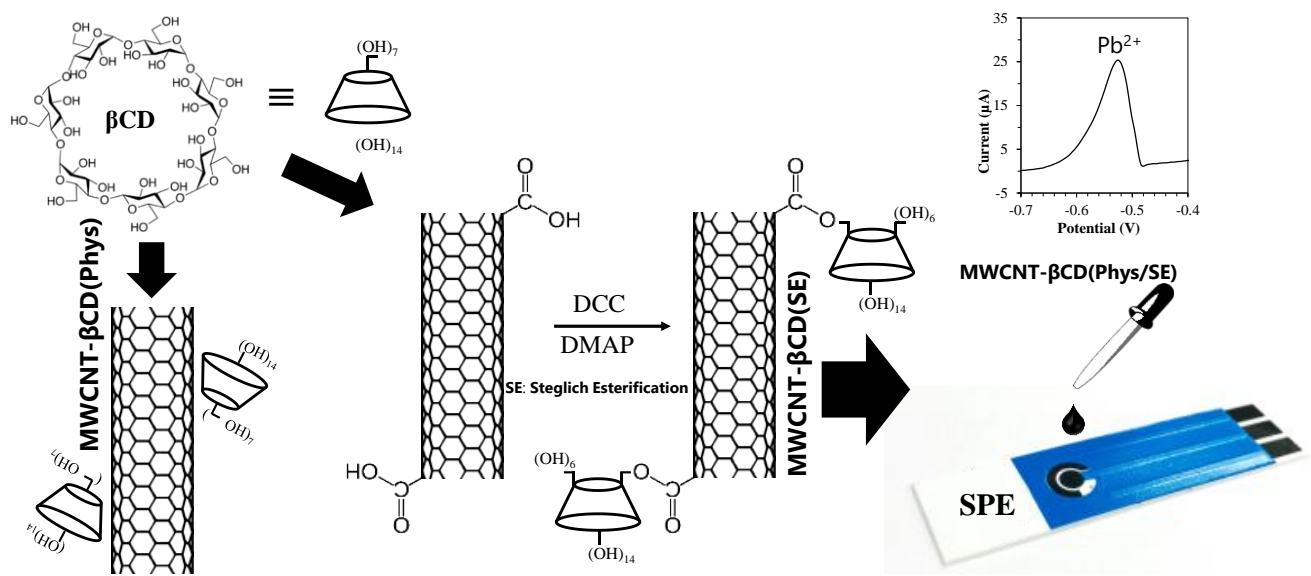

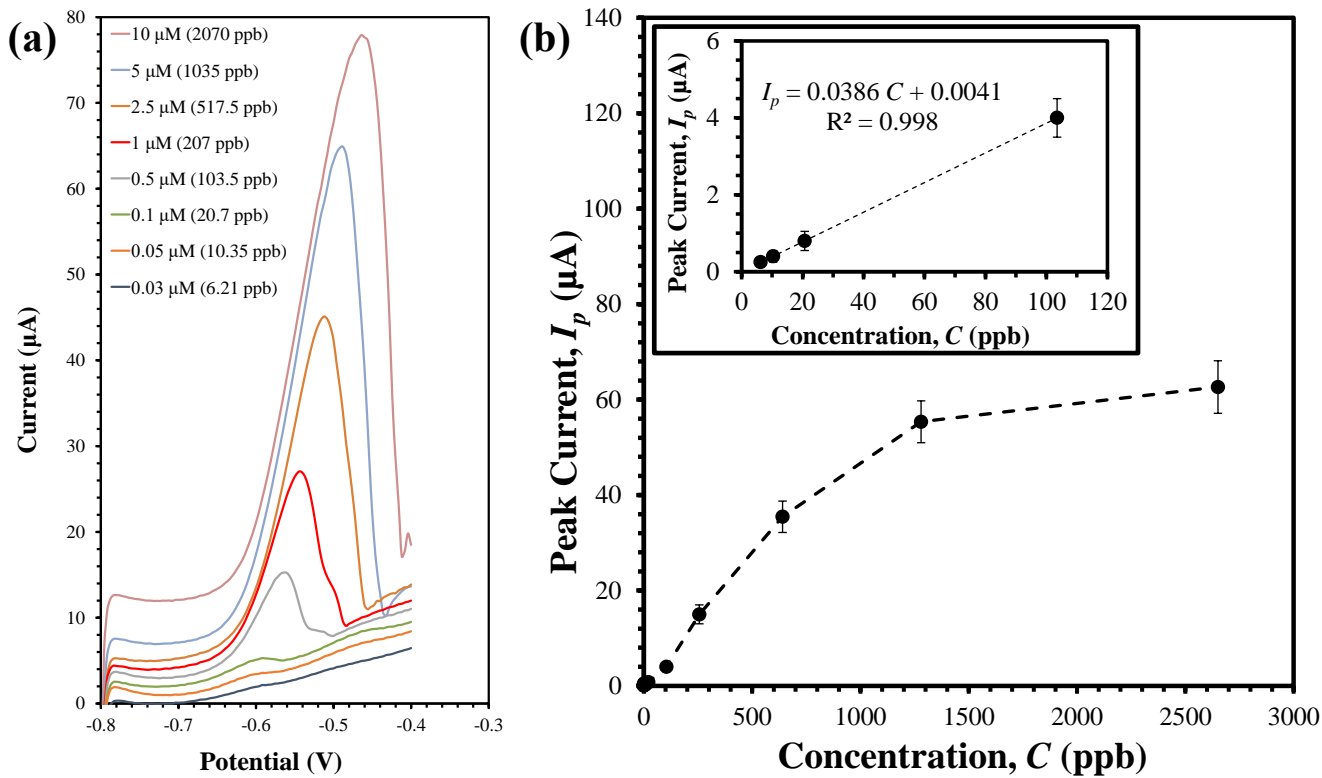

**Figure S1:** (a) DPASV of MWCNT-βCD (SE)/SPE in 0.1 M acetate buffer (pH 5) in the presence of 0.025−20 μM (6.2–103.3 ppb) Pb$^{2+}$. (b) Calibration curve (i.e., peak current vs Pb$^{2+}$ concentration) for MWCNT-βCD (Phys)/SPE. All of the experiments were done with a deposition potential of −0.8 V for 600 s under magnetic stirring before measurements.

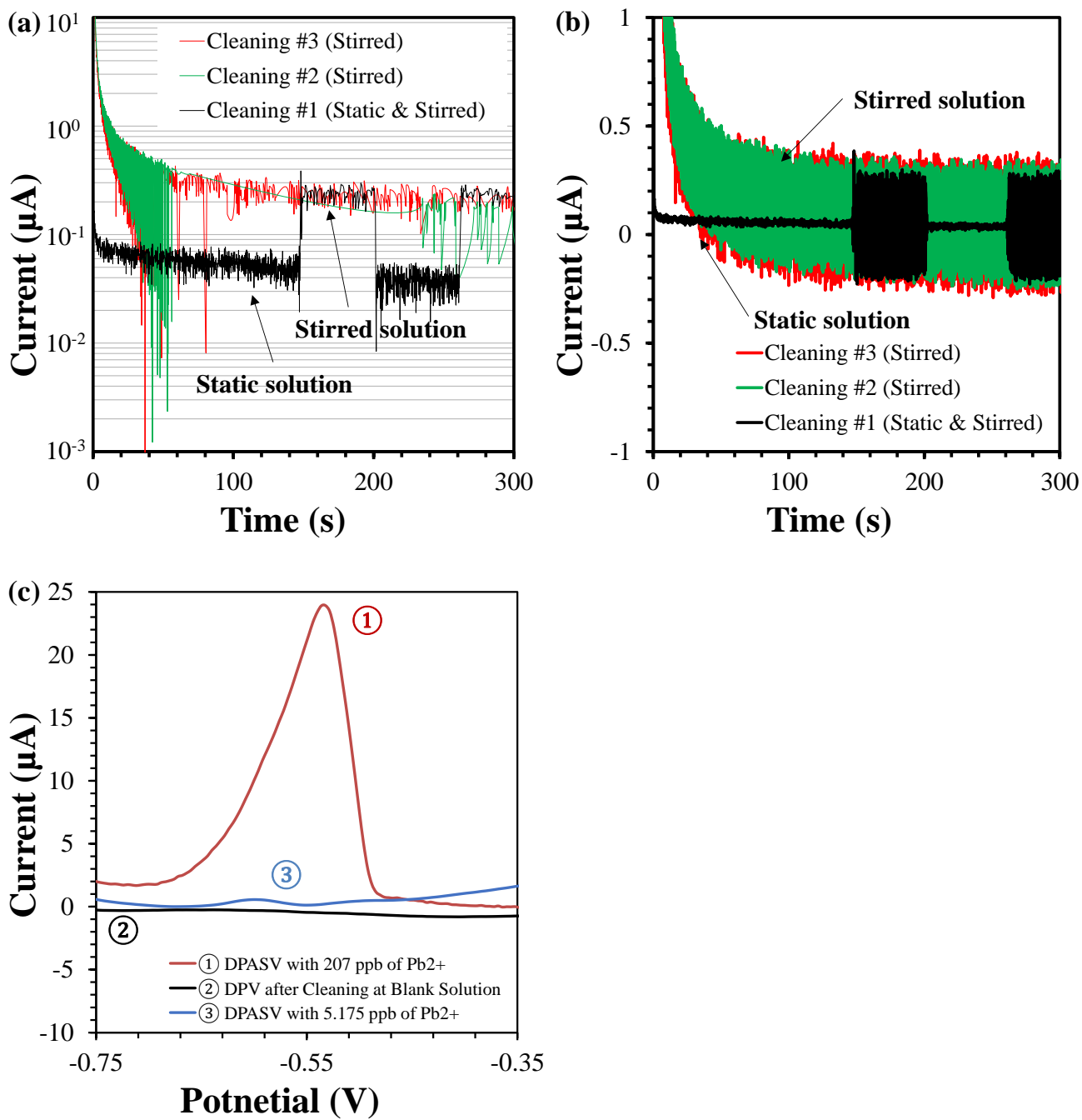

**Figure S2:** Electrochemical process of electrode surface cleaning/regeneration by applying a +0.31 V under magnetic stirring for 300 s. The resulting current response of the electrodes are shown with logarithmic (a) and (b) general numbers with respect to elapsed time. (c) The DPASV curves for new sensor (①), cleaned sensor (②) and reused sensor ③ for the detection of 207 ppb, blank and 5.2 ppb of $Pb^{2+}$.

**Table S1:** Comparison of the MWCNTs-based electrochemical sensors for Pb$^{2+}$

| Electrode | Modifications | Linear Range | Detection Limit | Comments | Year[Ref.] |
|---|---|---|---|---|---|
| MWCNT-GC | Poly(PCV)/Bi | 1-200 ppb | 0.4 ppb | MWCNT modified with poly(PCV) and Bi film | 2015[25] |
| MWCNT-GC | BiOCl/Nafion | 5-50 ppb | 0.57 ppb | bismuth-oxychloride particle-MWCNT composite | 2015[26] |
| MWCNTs/Bi-GC | -SH, -NH2, -COO-, -OH | 2-50 ppb | 0.3 ppb | N-doping and thiol-modification of MWCNT | 2016[27] |
| AuNPs-CNFs-GCE | Isopropanol/Nafion | 20.7-207.2 ppb | 20.7 ppb | AuNPs/CNFs fabricated via electrospinning and in situ thermal reduction | 2016[28] |
| CNT Thread | None | 207-1035 ppb | 0.12 ppb | electrochemical cell fabricated by CNT thread | 2017[29] |
| MWCNT-IIP/Pt | Ion Imprinted polymer | 1-5 ppm | 0.02 ppb | binding sites for lead ions sculpted with lead ion as template and NNMBA-crosslinked polyacrylamide as the solid matrix on MWCNTs | 2018[30] |
| MWCNT-βCD | Physical, Steglich esterification | 3.1-103.3 ppb | 0.9 ppb | Physically and Steglich esterification modified MWCNT and βCD | This work |

**Table S2:** Breakdown of materials cost to fabricate a single MWCNT-βCD (Phys/SE)/SPE electrode

| Name of Materials | Unit Price | Price (each) | Price (for one sensor) | Price for one experiment |
|---|---|---|---|---|
| MWCNTs | $125/25gm | $.005 (each mg) | 0.02 mg × $0.005 = $0.0001 [10 µL solution with 2 mg/mL concentration of MWCNT] | |
| βCD | $220/100gm | $0.0022 (each mg) | 0.02mg × $0.0022 = $0.00044 [10 µL solution with 2 mg/mL concentration of βCD] | |
| SPE | $50/40 pieces | $1.25 (each) | $1.25 (each) | |
| MWCNT-βCD | | | $2.64×10$^{-3}$<br><br>Cost of one drop of **MWCNT-βCD** with respect to one **SPE** = $2.64×10$^{-3}$ / $1.25 = 0.21% | $2.64×10$^{-4}$ [for reusing 10 times] |
| MWCNT-βCD /SPE | | | $1.25+$0.0001+$0.00044 = $1.25054 | $0.0250108 [for 50 measurements] |